# Highly Anisotropic Elastic Properties of Suspended Black Arsenic Nanoribbons

Yunfei Yu,[1,2] Guoshuai Du,[1,2] Shang Chen,[1] Jingjing Zhang,[1,2] Yubing Du,[1,2] Qinglin Xia,[3] Ke Jin,[1] and Yabin Chen[1,2,4,*]

[1]*Advanced Research Institute of Multidisciplinary Sciences (ARIMS), Beijing Institute of Technology, Beijing 100081, P. R. China*

[2]*School of Aerospace Engineering, Beijing Institute of Technology, Beijing 100081, P. R. China*

[3]*School of Physics and State Key Laboratory of Powder Metallurgy, Central South University, Changsha, Hunan 410083, P. R. China*

[4]*BIT Chongqing Institute of Microelectronics and Microsystems, Chongqing 400030, P. R. China*

[*]Correspondence and requests for materials should be addressed to: chyb0422@bit.edu.cn (Y.C.)

**ABSTRACT:** Anisotropy, as an exotic degree of freedom, enables us to discover the emergent two-dimensional (2D) layered nanomaterials with low in-plane symmetry and to explore their outstanding properties and promising applications. 2D black arsenic (b-As) with puckered structure has garnered increasing attention these years owing to its extreme anisotropy with respect to the electrical, thermal, and optical properties. However, the investigation on mechanical properties of 2D b-As is still lacking, despite much effort on theoretical simulations. Herein, we report the highly anisotropic elastic properties of suspended b-As nanoribbons via atomic force microscope-based nanoindentation. It was found that the extracted Young's modulus of b-As nanoribbons exhibits remarkable anisotropy, which approximates to 72.2±5.4 and 44.3±1.4 GPa along zigzag and armchair directions, respectively. The anisotropic ratio reaches up to ~1.6. We expect that these results could lay a solid foundation for the potential applications of 2D anisotropic nanomaterials in the next-generation nanomechanics and optoelectronics.

**Table of Contents Graphic**

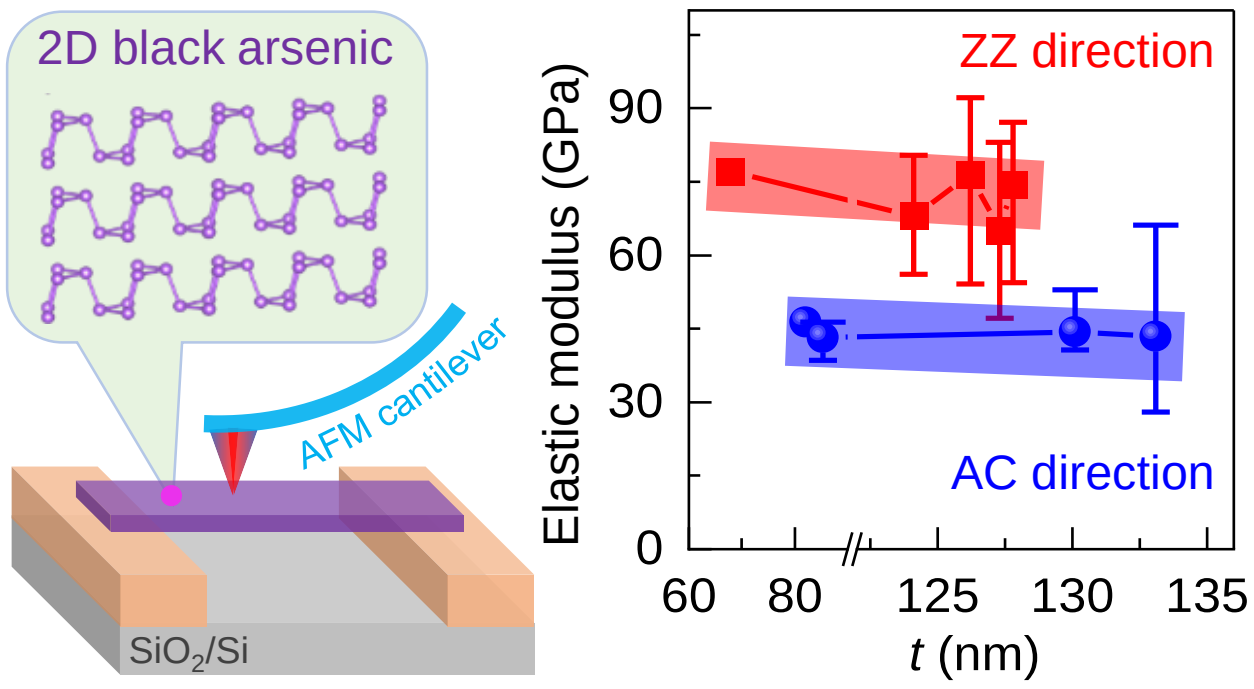

Anisotropic two-dimensional (2D) layered nanomaterials emerge these years and have attracted the extensive attention due to their unique atomic structures and distinguished physical properties, which intrinsically endow the orientation-dependent transport behaviors and numerous potential applications in nanodevices and nanoelectronics.[1-5] Intriguingly, anisotropic 2D nanomaterials generally possess the lattice structures with very low symmetry, much different from the isotropic 2D materials, including the well-studied six-fold graphene and transitional metal dichalcogenides.[6] As a new degree of freedom, anisotropy in specific 2D materials, such as the orthorhombic black phosphorus (b-P, space group cmce), render them two principle lattice axes in the basal plane, that is, the orthogonal zigzag (ZZ) and armchair (AC) directions.[7-9] Till now, many investigations have been performed to explore the anisotropic electrical, phonon vibrational, and thermal properties of 2D b-P.[10-12] For instance, the anisotropic ratio of thermal conductivity of b-P is determined to be ~1.8 in experiment, mainly attributed to the superior phonon velocity and scattering process along ZZ direction.[13]

As an extraordinary layered semiconductor, black arsenic (b-As) has been evidently proved to exhibit the extreme in-plane anisotropy of optical, thermal and electrical properties, originated from its similar puckered structure with b-P.[14] It is notable that field-effect transistor based on b-As still function well without the obvious degradation after exposing in ambient condition for ~26 days, implying its fantastic structural stability and application potentials.[15] Furthermore, 2D few-layer b-As displayed the asymmetric Rashba valleys and exotic quantum Hall states due to its centrosymmetric electronic properties and spin-orbit coupling effect.[16] Importantly, the systematic research on the anisotropic mechanical properties of 2D b-As, especially its lattice orientation-dependent behavior, is lacking till now, despite much effort on the theoretical calculations.[17-19]

Nanoindentation technique based on atomic force microscopy (AFM) is widely utilized to measure the elastic deformation and fracture behavior of numerous 2D layered nanomaterials, on the basis of suspended circular drum model or doubly clamped beam model.[20-24] It is found that the prominent elastic modulus and breaking strength of monolayer graphene reached as high as 1 TPa and 130 GPa, respectively, well consistent with the theoretical results.[21] More interestingly, 2D b-P exhibited the significant anisotropic mechanical properties, whereby the Young's modulus along ZZ direction is dramatically higher than that along the AC direction, confirmed by the combined experimental and theoretical results.[25-28] In this work, we attempted to measure the elastic modulus of b-As nanoflakes along different directions via AFM nanoindentation approach, and further unravel its extremely anisotropic mechanical properties. The suspended b-As nanoribbons along a specific lattice orientation with various thickness can be achieved by focused ion beam (FIB) process. We hope that the highly anisotropic elastic properties of b-As could shed light on 2D advanced nanomechanics and nanodevices.

Preparation of the suspended b-As nanoribbon is with great challenge. The b-As nanoflakes were obtained by mechanical exfoliation from bulk crystalline mineral, whereafter they were deterministically transferred onto Si/$SiO_2$ substrate with the pre-patterned arrays of "L" grooves by using all-dry viscoelastic stamping method.[29] The AC and ZZ directions of b-As nanoflakes were exactly determined by polarized Raman spectroscopy, as demonstrated in Figure 1a. Then, we employed FIB technique to shape the b-As nanoflakes along specific directions, resulting in the suspended b-As nanoribbons with the desired width/length ratio which are suitable for the following AFM indentation tests (Figures 1b and 1c). Obviously, the clean surface of b-As samples is free of any apparent cracks, wrinkles, or contaminations after the entire fabrication process. The array of suspended and narrow b-As nanoribbons along the orthogonal ZZ and AC directions allow us to systematically measure their mechanical properties. Figure 1d demonstrates the typical Raman spectrum of the exfoliated b-As nanoflake before FIB process, which displayed three characteristic phonon modes as the out-of-plane $A_g^1$, and in-plane $B_{2g}$ and $A_g^2$. To accurately identify the lattice orientation of b-As, angle-resolved polarized Raman spectroscopy (ARPRS) and temperature-dependent polarized Raman spectroscopy (TDPRS) were jointly explored.[30-33] Based on the identical lattice dynamics and group theory with b-P, the ARPRS results showed that Raman intensity $I(B_{2g})$ of $B_{2g}$ mode of b-As nanoflake varied as the polarization angle $\theta$ of excitation source, leading to the periodic polar plot with four-fold symmetry. As exhibited in Figure 1e, the obtained data about its Raman intensity versus polarization angle can be well fitted with $I(B_{2g}) \propto \cos^2 2\theta$, where the minimum values imply the AC or ZZ direction of b-As. Furthermore, we measured the temperature- dependent Raman spectrum of b-As, following TDPRS approach. Importantly, Raman shift of $A_g^2$ mode presents more significant temperature-dependence when the laser polarization is aligned with ZZ ($-0.0230$ $cm^{-1}$/K) than the counterpart AC ($-0.0199$ $cm^{-1}$/K) direction in Figure 1f (more data shown in Figure S1), mostly due to its anisotropic phonon velocity and thus superior thermal conductivity, which is well consistent with literature data.[32]

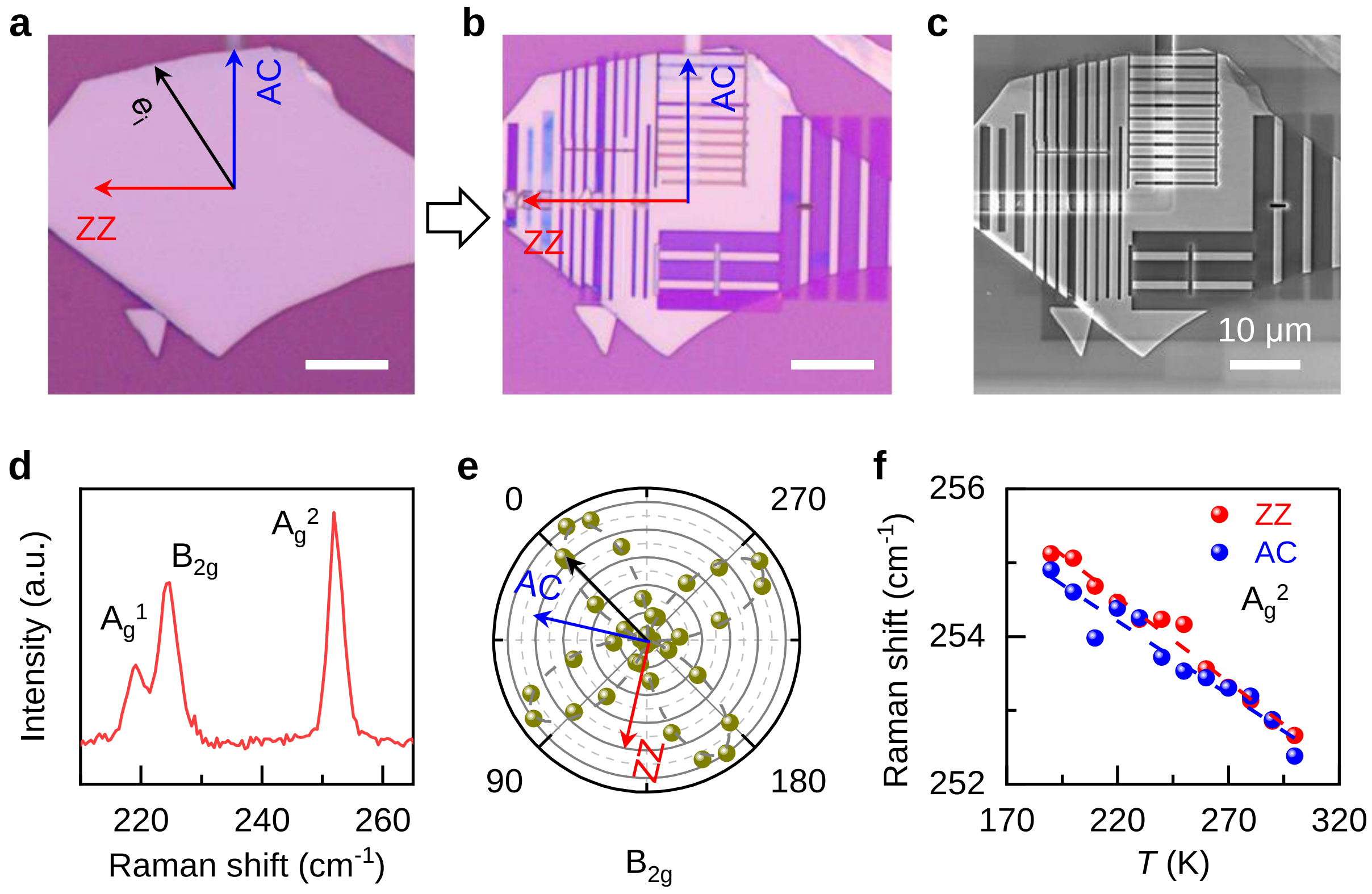

**Figure 1**. Structural characterizations and Raman spectrum of b-As nanoribbons. **a**) Optical image of the exfoliated b-As nanoflake before FIB shaping. The Si/$SiO_2$ substrate is with the pre-patterned arrays of "L" grooves. The ZZ (red line) and AC (blue line) directions are determined by using polarized Raman spectroscopy. **b**) Optical image of b-As nanoflake after FIB shaping. The fabricated b-As nanoribbons bridge the "L" grooves. The red and blue lines indicate the ZZ and AC directions of b-As nanoribbons, respectively. **c**) Scanning electron microscopy (SEM) image of the same b-As flake in (b) with high resolution. All scale bars are 10 μm. (**d**) The representative Raman spectrum of b-As nanoflake with the labeled $A_g^1$, $B_{2g}$ and $A_g^2$ modes. (**e**) ARPRS results of b-As sample in polar coordinates. The gray dashed line represents the fitting result of $B_{2g}$ mode with four-fold symmetry. (**f**) The Raman shift evolution of $A_g^2$ phonon mode as a function of temperature. The red and blue data points were measured with laser polarization along the ZZ and AC directions of b-As nanoribbon, respectively. The dashed lines are the linear fitting results.

Next, we turn to measure the anisotropic mechanical properties of the prepared b-As nanoribbons along different lattice orientations by using AFM nanoindentation method, and further attempt to extract the elastic modulus through the force-indentation depth curves. As shown in Figure 2a, it is obvious that b-As nanoribbons along ZZ direction was suspended on the trench of $SiO_2$/Si substrate. Typically, the width and depth of the pre-patterned "L" groove approximated to 2 μm and 250 nm, respectively, to weaken the test error of AFM indentation.[34] Thickness of this b-As nanoribbon was measured as 124.1 nm from the line profile of the magnified AFM image obtained through tapping mode in Figure S2. In order to exactly acquire elastic modulus, the b-As nanoribbon can be simplified as an elastic beam by following the double-clamped beam model,[34] and the variable load applied by AFM probe was concentrated at the center of the suspended portion (Figure 2b). Therefore, the controlled load $F$ is theoretically correlated with the elastic deformation $\delta$ of b-As nanoribbon before fracture behavior by the following equation[28] $F = \frac{\pi^4 Ewt^4}{6l^3}\left(\frac{\delta}{t}\right) + \frac{\pi^4 Ewt^4}{8l^3}\left(\frac{\delta}{t}\right)^3 + \frac{Tt}{l}\left(\frac{\delta}{t}\right)$, where $E$ and $T$ are the effective elastic modulus and pretension of b-As nanoribbon, respectively; $l$, $w$, and $t$ are the length, width, and thickness of the b-As nanoribbon, respectively. Notably, the normalized indentation depth to the sample thickness $\delta/t$ is dimensionless. The first and second terms correspond to the linear (especially in small deformation limit) and non-linear behaviors in the load force-displacement curves, respectively. The third one represents the initial pretension of b-As nanoribbon.

We first investigated the elasticity of b-As nanoribbon along the ZZ direction. Before nanoindentation test, the b-As nanoribbons were cyclically scanned until the thermal drift was negligible. As depicted in Figure 2c, it is manifest that the load-indentation depth relationship varied at the different deformation regions. Quantitatively, the load linearly increased with indentation depth at the beginning, followed by the evident cubic power of deformation. In principle, the larger $F$ - $\delta$ slope corresponds to the stronger tensile strength. The fitted curve under load ~7263 nN in Figure 2c reveals that the extracted Young's modulus was ~68.0 GPa for b-As nanoribbons (124.1 nm thick) along the ZZ direction, which is in good agreement with the result of density functional theory calculations.[18] Meanwhile, the cyclic $F$ versus $\delta/t$ curves were consecutively acquired with the various maximum loads ranging from 4060 to 7263 nN, as shown in Figure 2d. Importantly, their consistency and repeatability confirmed a quite stable and firm contact between b-As and $SiO_2$ substrate, and further reflects the acceptable accuracy of our AFM nanoindentation tests.

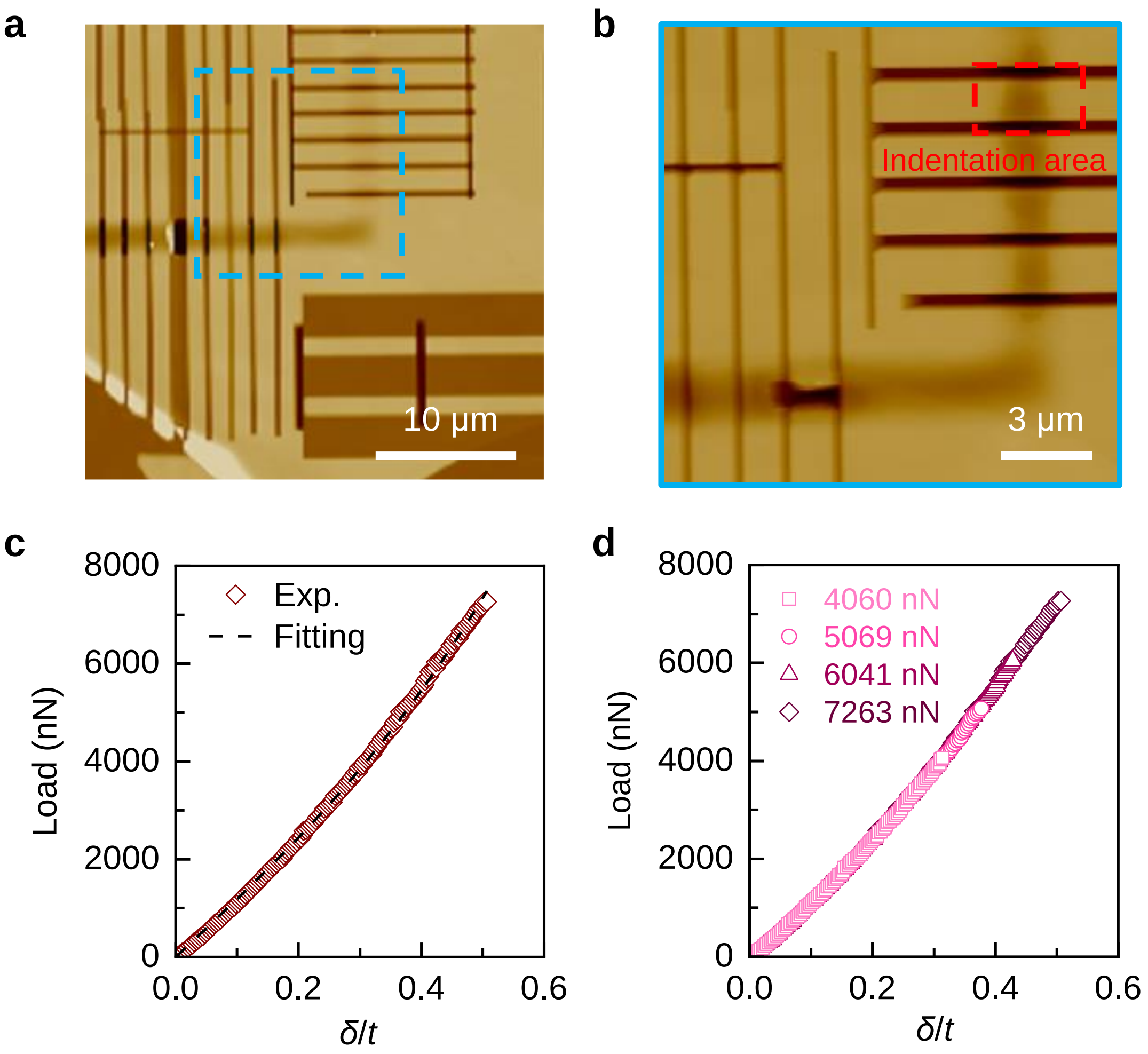

**Figure 2.** AFM nanoindentation test for elastic properties of b-As nanoribbon along the ZZ direction. **a**) AFM image of the b-As nanoflake after FIB shaping with many suspended nanoribbons on Si/$SiO_2$ substrate. **b**) The detailed topographic result of the b-As nanoribbons as indicated by the light-blue dashed square in (a). Red dashed rectangular corresponds to the region for AFM nanoindentation. The thickness is around 124.1 nm. **c**) The representative load-displacement curve of b-As nanoribbon along the ZZ direction. The maximum load reached 7263 nN. The black dashed line is the fitting result via double-clamped beam model. **d**) Four consecutive load-displacement curves of a given b-As nanoribbon under variable loading forces, ranging from 4060 to 7263 nN. The data are obviously consistent with each other.

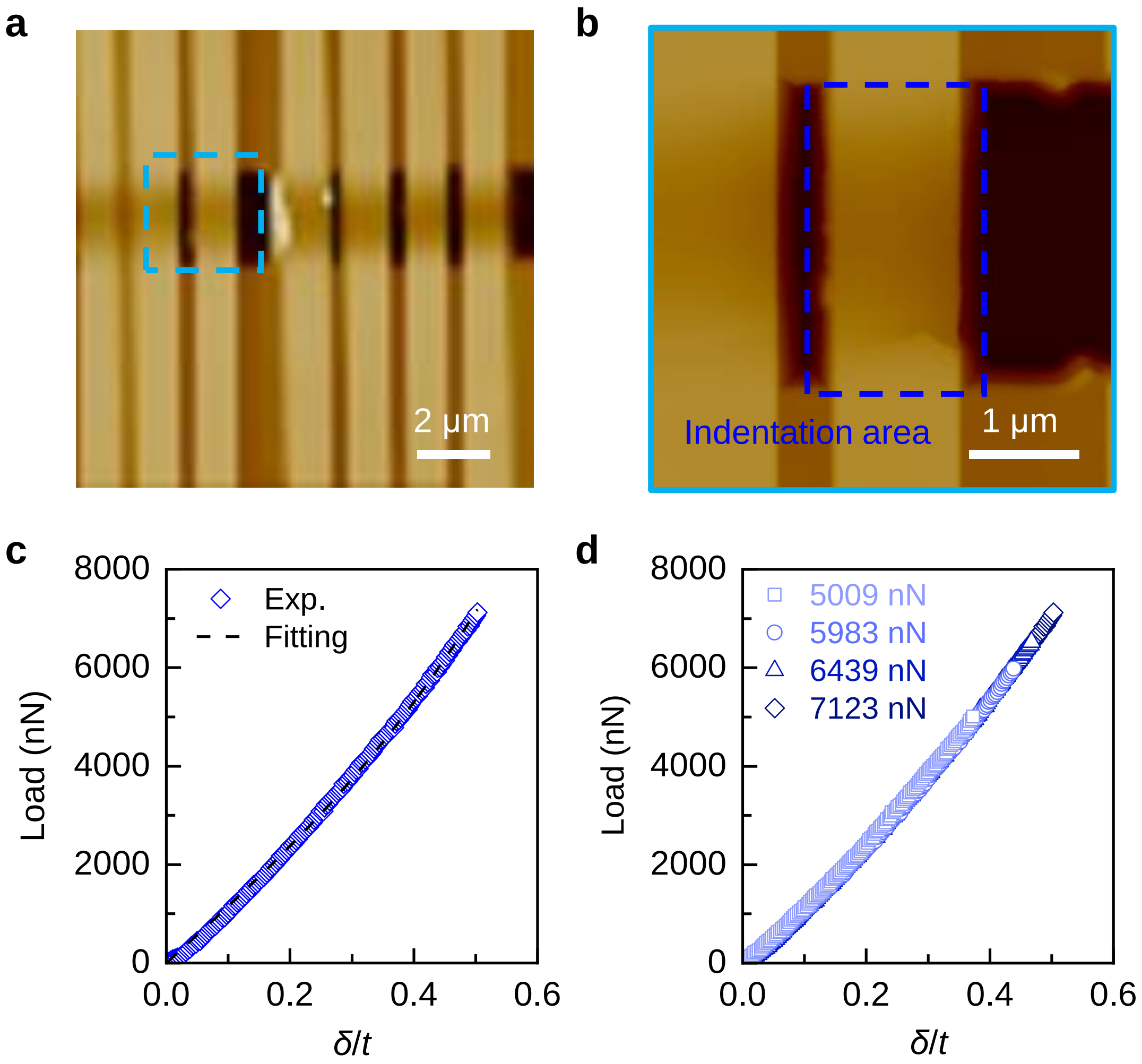

**Figure 3.** AFM nanoindentation for elastic properties of b-As nanoribbon along the AC direction. **a**) AFM image of b-As nanoribbons after FIB suspended on the pre-patterned grooves. **b**) Topographic result of the b-As nanoribbon highlighted with light-blue dashed square in (a). Blue dashed rectangular means the zone for AFM nanoindentation measurement. The thickness is determined to be 130.1 nm. **c**) The typical load-displacement curve of b-As nanoribbon along the AC direction. The black dashed line is the fitting result by using the equation as discussed above. The experimental points are acquired with the maximum load ~ 7123 nN. **d**) Four load-displacement curves of b-As nanoribbon along the AC direction under the different loading forces from 5009 to 7123nN.

In comparison with the effective modulus of b-As along the ZZ direction, we further characterized its elastic properties along the AC direction. In order to maximize the test precision, the same large and crystalline b-As nanoflake (Figure 2) was taken to fabricate b-As nanoribbons along the AC direction, as shown in Figure 3a. The results of AFM morphological image confirmed that its local thickness reached ~130.1 nm in Figures 3b and S2. Similarly, AFM tip precisely applied the controllable force concentrated at the center of the b-As beam. It is apparent that indentation load gradually increases till the maximum load ~7123 nN, from the initial linear to distinct non-linear dependence on depth (Figure 3c). Eventually, the fitted effective Young's modulus of b-As along the AC direction was determined to be ~44.4 GPa, which is dramatically weaker than that along the ZZ direction. The anisotropy of elastic modulus of b-As nanoflake reached 1.5. In addition, the consecutive nanoindentation tests were cyclically carried out with the variable loading forces, ranging from ~5000 to 7123 nN, and the results are well consistent with each other, as shown in Figure 3d.

Moreover, we tentatively elaborated the physical origin of the anisotropic modulus and its layer-dependence of b-As nanoflake. As illustrated in Figure 4a, b-As monolayer displayed the exotic puckered structure, in which each arsenic atom is covalently bonded with three other adjacent atoms. Ultimately, the non-coplanar arsenic atoms constitute the low symmetric lattice with the orthogonal ZZ and AC directions. In terms of nanomechanics, the larger lattice constant *b* along AC direction can expectably weaken its binding energy and further fracture strength, giving rise to the anisotropic mechanical characteristic. In comparison, b-As along ZZ direction can reasonably endure smaller elastic strain limit, which has been evidently proved in our recent studies.[35,36] Moreover, we tended to investigate the layer number-dependence of anisotropic mechanics by systematically measuring the Young's modulus of b-As nanoribbons with various thickness. As shown in Figure 4b, the obtained effective Young's modulus along the ZZ direction slightly decreases from ~77 to ~74 GPa when thickness ranged from 67.8 to 127.8 nm (more details shown in Figures S3 and S4). Meanwhile, the elastic modulus along the AC direction presented a negative thickness-dependence as well, primarily owing to the interlayer coupling for the thicker flakes. Clearly, the overall Young's modulus along the ZZ direction is remarkably stronger than that along AC direction, regardless of the b-As thickness in Figure 4b. The average anisotropic ratio of elastic modulus reached ~1.6, well consistent with the theoretical prediction.[37] To be honest, we cannot perform the AFM nanoindentation on few layer or even monolayer b-As nanoribbon, due to the commonly known challenges during sample preparation. Ideally, their mechanical moduli should be enhanced dramatically.

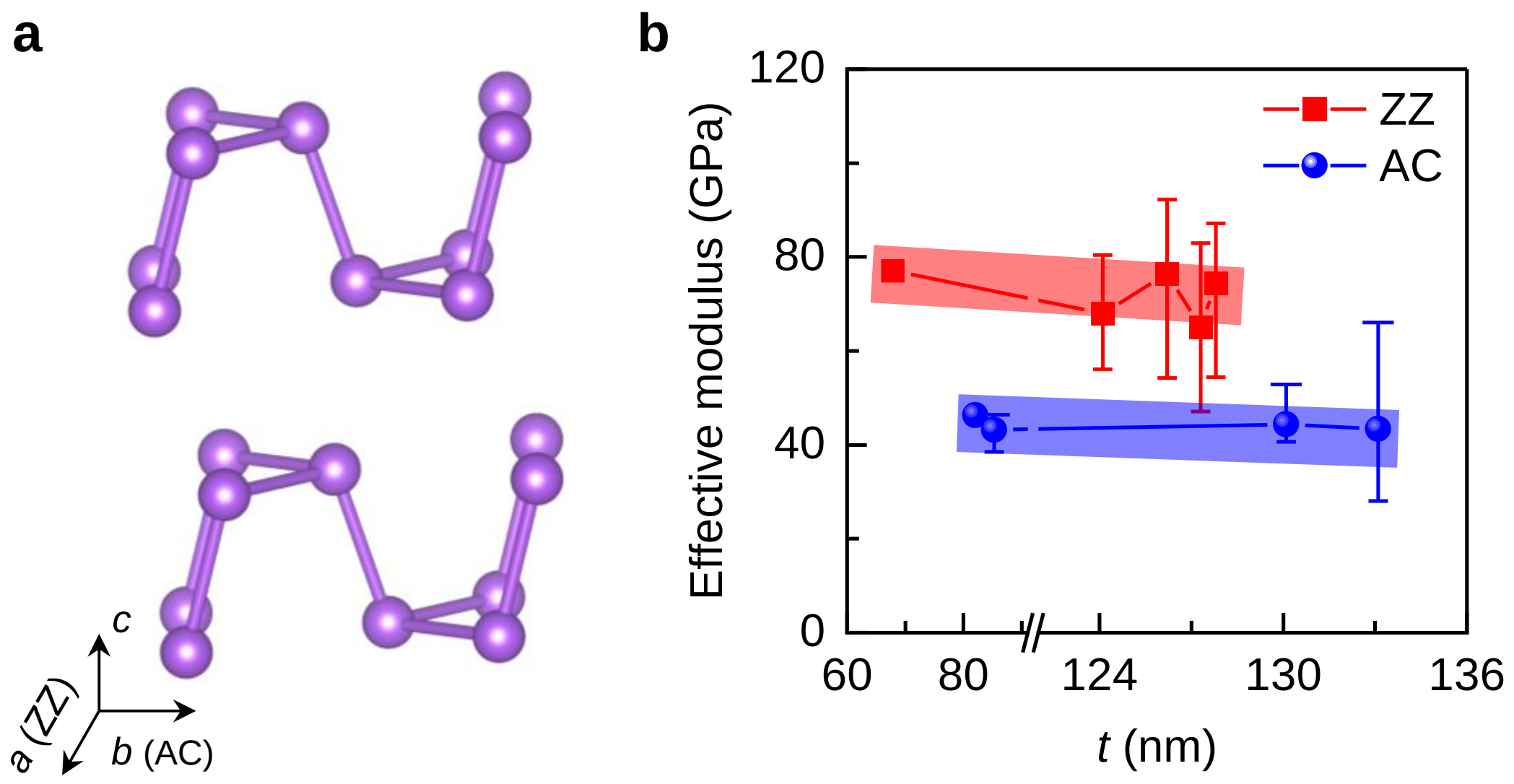

**Figure 4.** Thickness-dependence of Young's modulus of b-As nanoribbons along the ZZ and AC directions. **a**) The puckered lattice structure of the anisotropic 2D b-As. **b**) Thickness-dependence of effective modulus of b-As nanoribbons along the ZZ (red) and AC (blue) directions. The red and blue lines through the data points are guide to the eye. The obtained effective modulus along the ZZ direction is significantly stronger than that along the AC direction.

In summary, we demonstrated the various AFM nanoindentation measurements of the suspended b-As nanoribbons along the principle AC and ZZ directions. The lattice orientations of b-As nanoflakes were successfully identified by combing the ARPRS and TDPRS approaches. The suspended b-As nanoribbons have been realized by dry-transfer method and FIB process, allowing us to test its elastic behaviors along one specific direction. It was found that mechanical properties of b-As presented the extreme anisotropy. The superior Young's modulus of b-As along ZZ direction is dramatically larger than that along AC directions, owing to its puckered structures, and the remarkable anisotropic ratio was up to 1.6. We believe that this result could constitute a significant step forward to understand the anisotropic properties of 2D layered materials, and further boost many potential applications in nanodevices and nanoelectromechanical systems.

## EXPERIMENTAL METHODS

### Preparation and Characterization of Suspended b-As Nanoribbons

The b-As nanoflakes were prepared by mechanical exfoliation using polydimethylsiloxane instead of Scotch tape, in order to exclusive any contamination or tape residual. The original b-As crystal was from natural minerals. Then, the b-As nanoflakes were deterministically transferred onto Si/$SiO_2$ surface with controlled locations. The Si/$SiO_2$ substrate was pre-patterned with "L" groove arrays through all-dry viscoelastic stamping approach.[29]

To exactly identify lattice orientation of the b-As nanoflake, ARPRS and TDPRD measurements were carried out (Figure S1 in Supporting Information). The wavelength of laser excitation was 632.8 nm, and the laser power was as low as tens of μW to avoid the potential over-heating effect. In ARPRS experiment, b-As nanoflake was rotated from 0 to 180$^o$ with 10$^o$ interval. Angle-dependent Raman intensity of $B_{2g}$ mode can suggest the AC or ZZ direction of b-As. TDPRS measurements were further performed to exactly identify the AC and ZZ directions of b-As nanoflake. The laser polarization direction was parallel with one of the principle axial directions of b-As nanoflakes, and temperature varied from 180 to 300 K with 10$^o$ interval.

Thereafter, the b-As nanoflakes were patterned into nanoribbons with FEI Helios G4 UC dual beam system (SEM-FIB, Thermo Fisher Scientific). The acceleration voltage of the FIB process was 10 kV, and the beam current was 0.8 nA. In addition, b-As nanoflakes with different thickness were achieved by the direct exfoliation together with reactive ion etching technique (NRP-4000, NANO-MASTER), as shown in Figure S5. The optimized recipe was as follows: $O_2$:$CHF_3$ flow~5:45 sccm and chamber pressure ~170 mTorr. The crystalline quality of the b-As nanoflakes was preliminarily characterized by optical microscopy (BX53, Olympus), and further by Raman spectroscopy equipped with the Horiba iHR550 spectrometer. Thickness of b-As nanoflake was accurately determined by AFM (Dimension Icon, Bruker).

### AFM Nanoindentation

For AFM nanoindentation tests, the applied load by AFM tip was localized at the center of b-As nanoribbon. The load-indentation depth curve was acquired via PF-QNM mode during AFM measurements. In principle, spring constant $k$ of the silicon cantilever of each AFM tip can be precisely calibrated by the well-known Sader method in Figure S6, which is based on the harmonic oscillation model.[38] In our case, the representative spring constant and resonant frequency of AFM tip were 200 N/m and 525 kHz, respectively. Before recording the load-indentation depth data, each b-As nanoribbon was continuously scanned for multiple times till thermal drift can be negligible. Importantly, AFM morphology results did not show any evident slippage of b-As from the $SiO_2$ substrate. For the nanoindentation test, *z*-piezo displacement speed was tuned around 300 nm/s. The indentation depth $\delta$ was defined based on the central displacement of the suspended b-As nanoribbons. The applied load $F$ can be calculated through Hooke's law $F = kx$, where $x$ was the deflection of the probe cantilever and directly given by the AFM system. Therefore, indentation depth $\delta$ of the suspended sample can be derived as $\delta = z - x$, where $z$ is the vertical distance moved between AFM probe and b-As and can be directly offered by the AFM scanner.

## ASSOCIATED CONTENTS

### Supporting Information

The Supporting Information is available in the online version of this article.

Raman spectrum of 2D b-As samples before and after FIB processing; thickness measurement of b-As by AFM; SEM images and load-displacement data of b-As nanoribbons with different thickness; the typical results of b-As nanoflakes thinned by RIE process; characterizations of AFM probe used for nanoindentation.

### Conflict of Interests

The authors declare no competing financial interests.

### Data Availability

The data that support the findings of this study are available from the corresponding author upon reasonable request.

### Author Contributions

Y.C. conceived this research project and designed the experiment. Y.Y., J.Z., G.D., and Q.X. prepared the b-As samples. Y.Y., G.D., and Y.D. performed the AFM and Raman characterizations. Y.Y., and S.C. carried out the FIB cutting process for b-As nanoribbons, advised by K.J., and Y.C. Y.Y., and Y.C. wrote the manuscript with the essential input of other authors. All authors have given approval of the final manuscript.

### Acknowledgements

This work was financially supported by the National Natural Science Foundation of China (grant numbers 52072032, and 12090031), and the 173 JCJQ program (grant number 2021-JCJQ-JJ-0159). Prof. Q.L. Xia acknowledges the funding support by state key laboratory of powder metallurgy in Central South University.

***Supplementary Information:***

# Highly Anisotropic Elastic Properties of Suspended Black Arsenic Nanoribbons

Yunfei Yu,[1,2] Guoshuai Du,[1,2] Shang Chen,[1] Jingjing Zhang,[1,2] Yubing Du,[1,2] Qinglin Xia,[3] Ke Jin,[1] and Yabin Chen[1,2,4,*]

[1]*Advanced Research Institute of Multidisciplinary Sciences (ARIMS), Beijing Institute of Technology, Beijing 100081, P. R. China*

[2]*School of Aerospace Engineering, Beijing Institute of Technology, Beijing 100081, P. R. China*

[3]*School of Physics and State Key Laboratory of Powder Metallurgy, Central South University, Changsha, Hunan 410083, P. R. China*

[4]*BIT Chongqing Institute of Microelectronics and Microsystems, Chongqing 400030, P. R. China*

[*]Correspondence and requests for materials should be addressed to: chyb0422@bit.edu.cn (Y.C.)

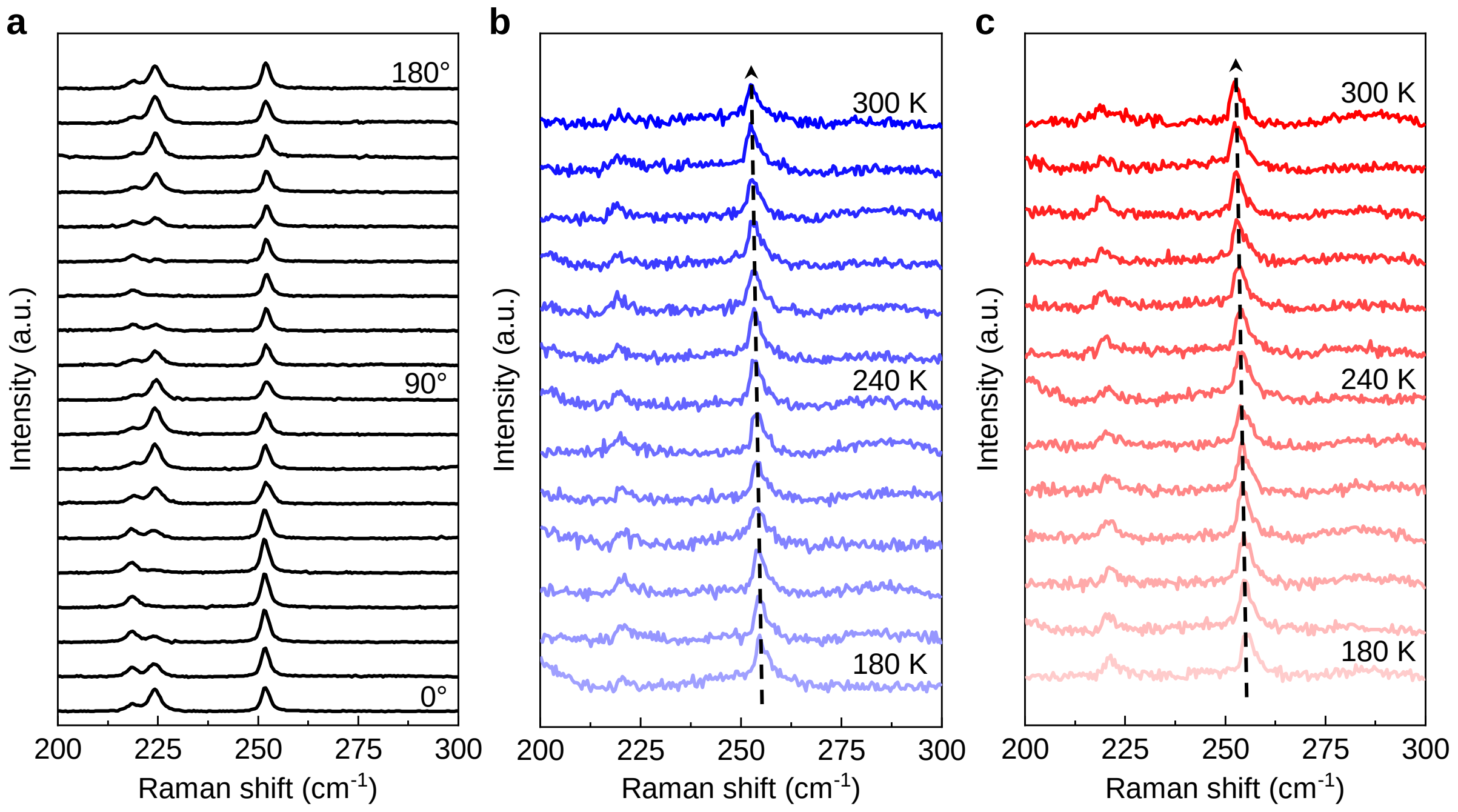

**Figure S1.** Angle-resolved polarized Raman spectrum of 2D b-As nanoflake and temperature-dependent polarized Raman spectrum of 2D b-As nanoribbons after FIB processing. (**a**) Angle-resolved polarized Raman spectrum of 2D b-As nanoflake at room temperature, ranging from $0^o$ to $180^o$ with $10^o$ interval. (**b-c**) Temperature-dependent polarized Raman spectrum of 2D b-As nanoribbons after FIB, and laser polarization aligned with AC (b) and ZZ (c) directions, respectively. Temperature varied from 180 to 300 K with 10 K interval. The laser power was low enough to avoid the potential overheating effect on 2D b-As samples.

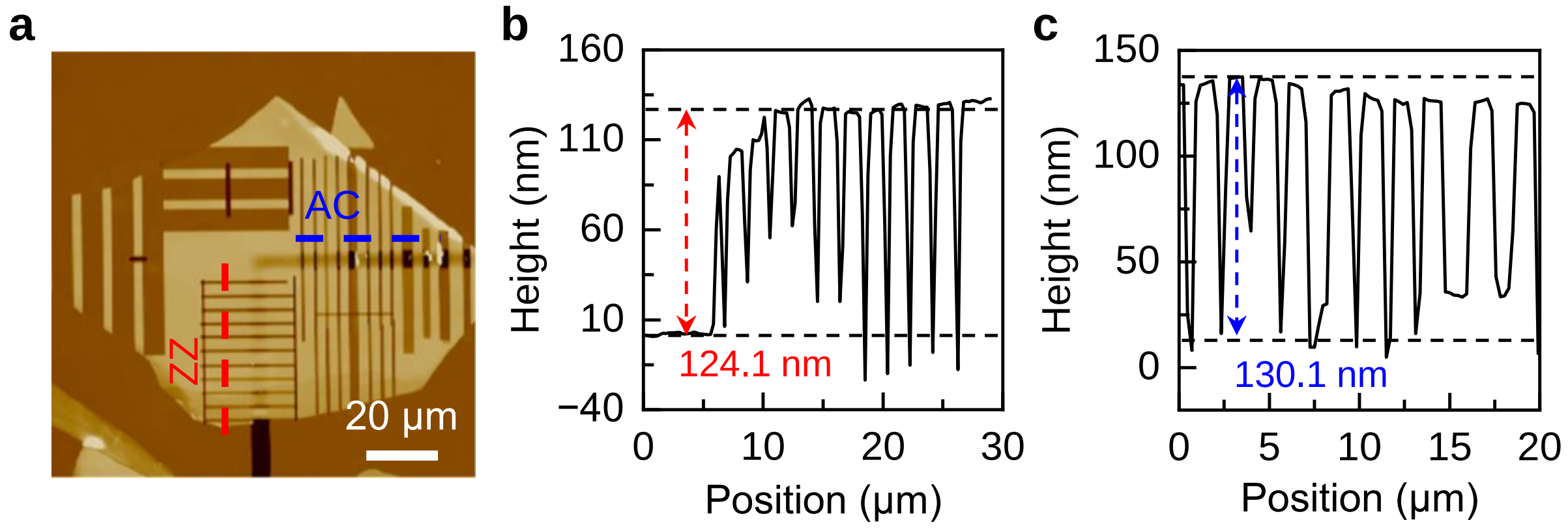

**Figure S2.** The representative AFM results used to determine the b-As thickness. (**a**) AFM image of 2D b-As nanoflake after FIB cutting. The red and blue dashed lines represent ZZ and AC directions, respectively. (**b**) Line profile for thickness measurement of b-As nanoribbon. The thickness was extracted as 124.1 nm. (**c**) Thickness measurement of b-As nanoribbons by AFM. The determined thickness was 130.1 nm.

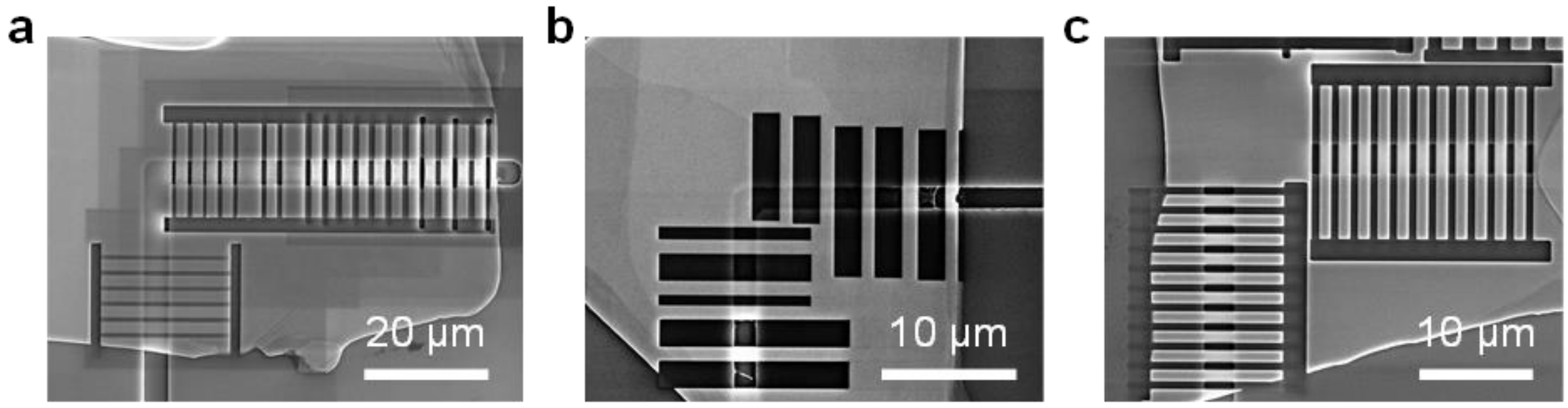

**Figure S3.** SEM images of the prepared 2D b-As nanoribbons with various thickness through FIB cutting. (**a-c**) The thickness of 2D b-As nanoribbons was determined to be around 97.4 (a), 118.8 (b), and 179.6 (c) nm.

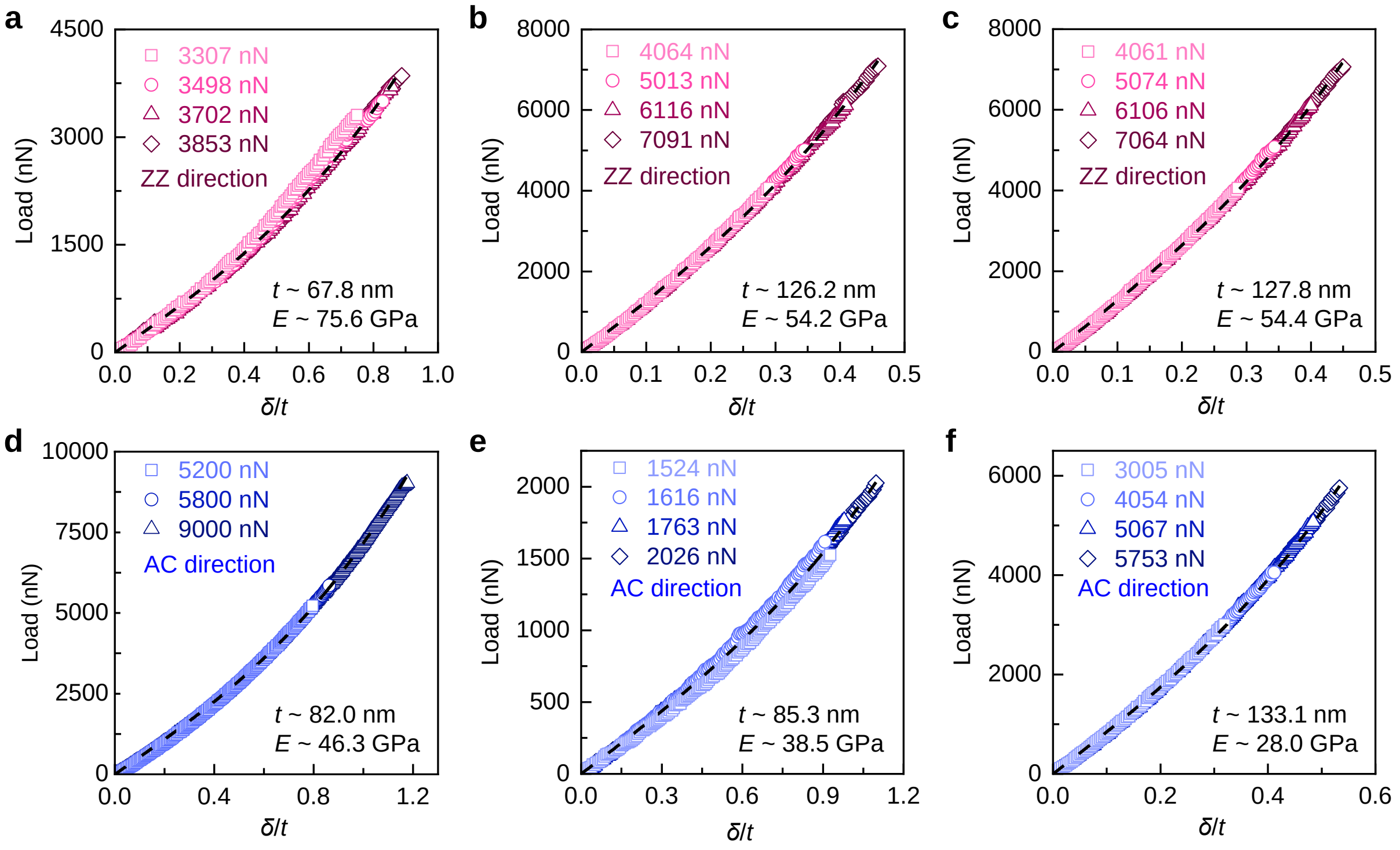

**Figure S4.** More loading force-displacement results of 2D b-As nanoribbons with various thickness. Black dashed lines are the fitting curve using the equation as discussed in the main text. (**a-c**) Experimental results of 2D b-As nanoribbons along the ZZ direction. The extracted modulus of these nanoribbons is 75.6, 54.2, and 54.4 GPa for 67.8 (a), 126.2 (b), and 127.8 (c) nm-thick nanoflakes, respectively. (**d-f**) Experimental results of 2D b-As nanoribbons along the AC direction. The sample thickness was 82.0 (d), 85.3 (e), and 133.1 (f) nm with the fitted elastic modulus of 46.3, 38.5, and 28.0 GPa, respectively.

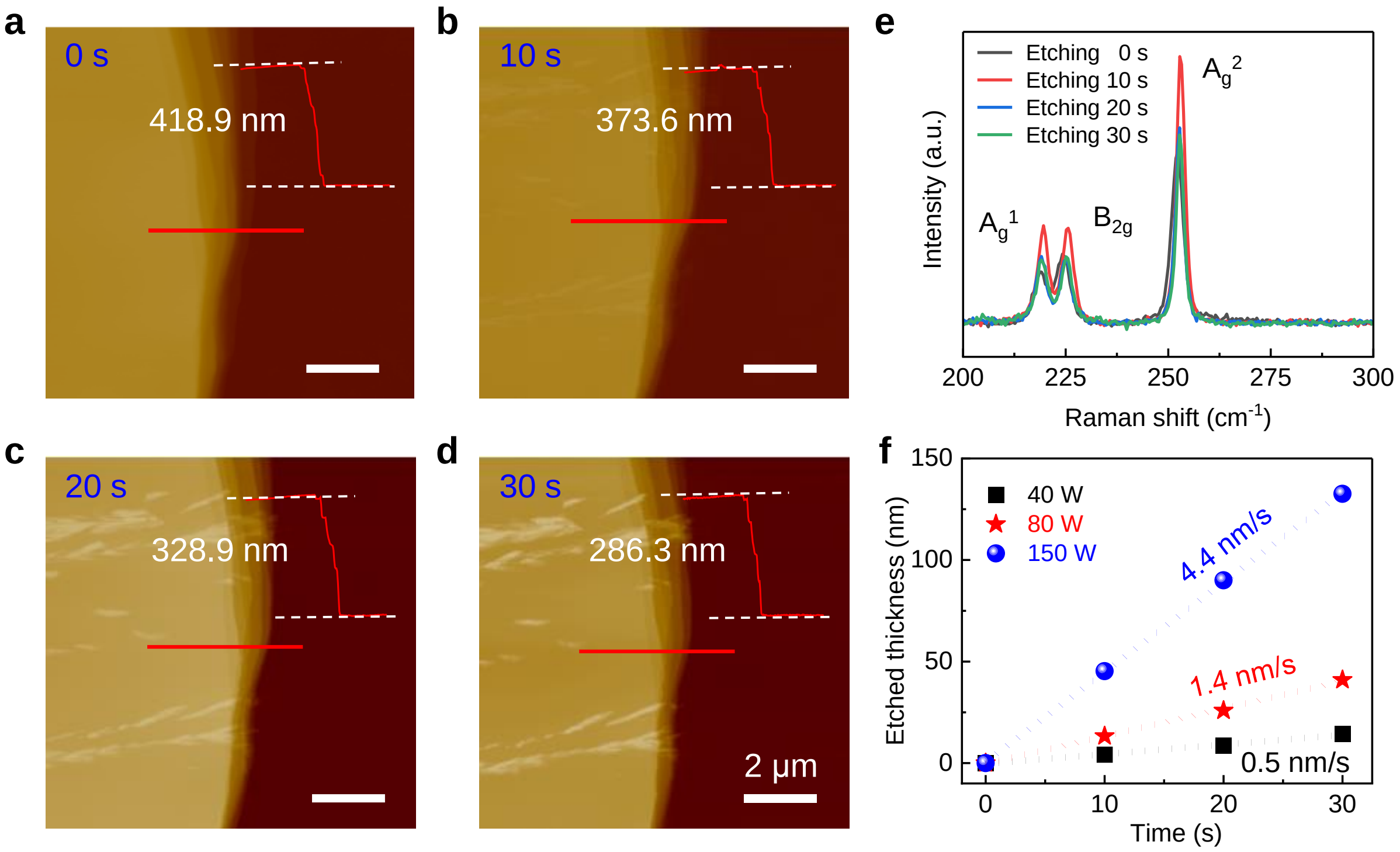

**Figure S5.** The typical results of b-As nanoflake thinned by RIE process. (**a-d**) AFM images of a given b-As nanoflake thinned with different etching period. The original thickness 418.9 nm (a) was reduced to 373.6 (b), 328.9 (c), and 286.3 (d) nm, after 10, 20, and 30 s etching, respectively. Notably, all thickness was measured at the same location (red line) of 2D b-As nanoflake. All scale bars are 2 μm. (**e**) Raman spectrum of b-As nanoflake after different etching times. Importantly, Raman intensity remains significantly strong, comparable with that before etching. (**f**) Etching rate of b-As nanoflake with different RF power. The linear slope (etching rate) was fitted as 0.5 (black), 1.4 (red), and 4.4 (blue) nm/s for 40, 80, and 150 W RF power.

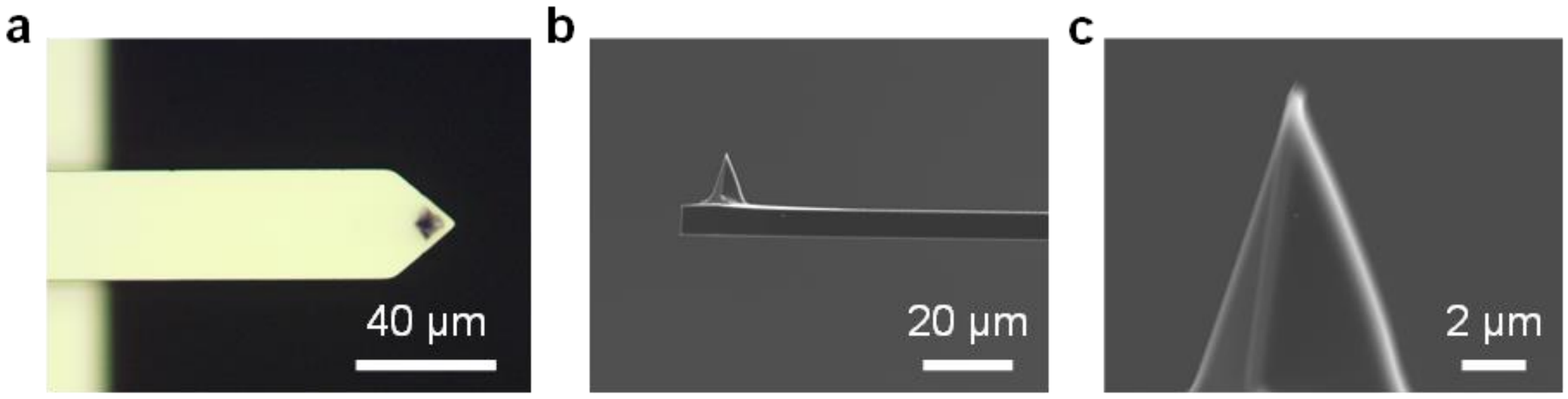

**Figure S6.** Characterization of AFM probe used for nanoindentation tests. (**a**) Optical image of the AFM probe. (**b-c**) SEM images of AFM tips with low (b) and high (c) magnifications.

The effective spring constant $k$ of the AFM probe is defined as $k = F / x$, and it can be principally determined by Sader method,[1] which states:

$$k = 0.1906 \rho b^2 L Q_f \omega_f \Gamma_i^f \left( \omega_f \right)$$

where $L$ and $b$ are the length and width of cantilever, respectively, which were measured directly by SEM system; $\rho$ is the density of fluid (air at 100 kPa and 298 K in our case). $\omega_f$ and $Q_f$ are the resonant frequency and quality factor of the fundamental resonance peak, respectively, which can be determined following simple harmonic oscillator in AFM system. $\Gamma_i^f \left( \omega_f \right)$ is the imaginary part of hydrodynamic function. For a quick calibration of $k$, the online calibration procedure is very useful (*https://ampc.ms.unimelb.edu.au/afm/webapp.html*), as suggested by the AFM manufacture and community.